\documentclass[runningheads]{llncs}
\usepackage[T1]{fontenc}
\usepackage{graphicx}
\begin{document}
\title{GastroVision: A Multi-class Endoscopy Image Dataset for Computer Aided Gastrointestinal Disease Detection}

\titlerunning{GastroVision}

\author{Debesh Jha$^*$\inst{1} \and
Vanshali Sharma$^*$\inst{2} \and
Neethi Dasu \inst{3}\and  
Nikhil Kumar Tomar\inst{1} \and
Steven Hicks\inst{4} \and
M.K. Bhuyan\inst{2} \and
Pradip K. Das\inst{2} \and
Michael A. Riegler\inst{4} \and
P{\aa}l Halvorsen\inst{4} \and
Ulas Bagci$\dag$\inst{1}  \and
Thomas de Lange$\dag$\inst{5}
}
\def\thefootnote{*}\footnotetext{These authors contributed equally to this work.}
\def\thefootnote{$\dag$}\footnotetext{Shared senior authorship.}

\authorrunning{D.Jha et al.}
\institute{Department of Radiology, Northwestern University, Chicago, USA \\
\email{\{debesh.jha, nikhil.tomar, ulas.bagci\}@northwestern.edu}\\ \and
Indian Institute of Technology Guwahati, Assam, India \\
\email{\{vanshalisharma, mkb, pkdas\}@iitg.ac.in} \\ \and
Department of Gastroenterology, Jefferson Health NJ, Cherry Hill, USA \\
\email{nrdasu4@gmail.com } \\ \and
SimulaMet, Oslo, Norway \\
\email{\{steven, michael, paalh\}@simula.no}\and
Department of Medicine and Emergencies - Mölndal Sahlgrenska University Hospital, Region Västra Götaland, Sweden  \& Department of Molecular and Clinical Medicin, Sahlgrenska Academy, University of Gothenburg, Sweden \\
\email{thomas.de.lange@gu.se }}
\maketitle              

\begin{abstract}
Integrating real-time artificial intelligence (AI) systems in clinical practices faces challenges such as scalability and acceptance. These challenges include data availability, biased outcomes, data quality, lack of transparency, and underperformance on unseen datasets from different distributions. The scarcity of large-scale, precisely labeled, and diverse datasets are the major challenge for clinical integration. This scarcity is also due to the legal restrictions and extensive manual efforts required for accurate annotations from clinicians. To address these challenges, we present \textit{GastroVision}, a multi-center open-access gastrointestinal (GI) endoscopy dataset that includes different anatomical landmarks, pathological abnormalities, polyp removal cases and normal findings (a total of 27 classes) from the GI tract. The dataset comprises 8,000 images acquired from  B{\ae}rum Hospital in Norway and Karolinska University Hospital in Sweden and was annotated and verified by experienced GI endoscopists. Furthermore, we validate the significance of our dataset with extensive benchmarking based on the popular deep learning based baseline models. We believe our dataset can facilitate the development of AI-based algorithms for GI disease detection and classification. Our dataset is available at \url{https://osf.io/84e7f/}. 

\end{abstract}
\section{Introduction}
\label{introduction}
Gastrointestinal (GI) cancers account for 26\% of cancer incidences and 35\% of cancer-related deaths worldwide. In 2018, there were approximately 4.8 million new cases of GI cancer and 3.4 million deaths~\cite{arnold2020global}. The five major types of GI cancers are colo-rectal (1.93 million cases; third most common cancer), pancreas (466,003 deaths; lowest survival rate), liver (905,677 cases), stomach (1.09 million cases), and esophagus (604,100 cases)~\cite{globalcancerobservatory}. These cancer cases are predicted to increase by 58\%, and related deaths could show a 73\% rise by 2040~\cite{arnold2020global}. Early detection of such cancers and their precursors can play an important role in improving the outcome and make the treatment less invasive. Some of the common examinations performed for GI cancer detection include endoscopy or esophagogastroduodenoscopy (EGD), capsule endoscopy, colonoscopy, imaging studies (MRI, X-ray, ultrasound, CT scan, or PET scan) or endoscopic ultrasound (EUS). Endoscopy is widely accepted as the gold standard for detecting abnormalities of the bowel lumen and mucosa,  upper endoscopy for esophagus, stomach, and the duodenum and  colonoscopy for the large bowel and rectum GI tract for abnormalities, respectively.

The endoscopies are performed by nurses or doctor endoscopists. The assessment of the endoscopy examinations is operator dependent, and the assessment and therapeutic decision vary between endoscopists. Consequently, the quality and accuracy of detection and diagnosis of lesions are attributed to the level of the operator skills and efforts of the endoscopists. Despite various measures taken to provide guidance for operators, significant lesion miss rates are still reported. For example, there is evidence of colon polyp miss rates of up to 27\% due to polyps and operator characteristics~\cite{ahn2012miss,mahmud2015computer}. Considering the shortcomings of the manual review process, various automated systems are adopted to provide AI-based real-time support to clinicians to reduce lesion miss rates and misinterpretation of lesions to ultimately increase detection rates. A microsimulation study reports a 4.8\% incremental gain in the reduction of colorectal cancer incidence when colonoscopy screening was combined with AI tools~\cite{areia2022cost}. Such findings motivated research work in healthcare to adopt AI as a potential tool for GI cancer detection. Gastric cancer, inflammatory bowel disease (IBD), and esophageal neoplasia are some of the GI tract findings already being investigated using AI techniques~\cite{abadir2020artificial}. Despite AI being adopted in some hospitals for clinical applications, the integration of AI into the extensive clinical setting is still limited. Integrating AI techniques with regular clinical practices is multifactorial and poses serious concerns regarding its implementation in large-scale healthcare systems. One of the significant factors is the algorithmic bias, which worsens when the system learns from the annotations handled by a single, non-blinded endoscopist who may have personal thresholds to label the findings. 

Moreover, most existing AI models depend on data acquired from a single center, which makes them less valid when faced with a varied patient population. This leads to spectrum bias under which AI systems encounter performance drops due to the significant shift in the original clinical context and the test sample population. In such cases, unexpected outcomes and diagnostic accuracy could be obtained using automated tools. Such bias issues could reach the clinical systems at any point of the process, including data collection, study design, data entry and pre-processing, algorithm design, and implementation. The very beginning of the process, i.e., data collection, is of utmost importance for reproducibility and to perform validations on images from a diverse population, different centers, and imaging modalities. To develop scalable healthcare systems, it is vital to consider these challenges and perform real-time validations. However, the scarcity of comprehensive data covering a range of real-time imaging scenarios arising during endoscopy or colonoscopy makes it difficult to develop a robust AI-based model. Although much progress has been made on automated cancer detection and classification~\cite{jha2021real,li2021colonoscopy}, it is still challenging to adapt such models into real-time clinical settings as they are tested on small-sized datasets with limited classes.

Some classes in the dataset could be scarce because some conditions or diseases occur less often. Consequently, such findings are not frequently captured and remain unexplored despite requiring medical attention. AI-based detection of these findings, even with a small sample count, can significantly benefit from techniques like one-shot or few-shot learning. These techniques allow the AI models to learn patterns and features indicative of the condition, thus, enabling accurate diagnosis with minimal training data. Apart from the above-mentioned limitations, many existing datasets are available on request, and prior consenting is required, which delays the process and does not guarantee accessibility. Therefore, in this paper, we publish \textit{GastroVision}, an open-access multi-class endoscopy image dataset for automated GI disease detection that does not require prior consenting and can be downloaded easily with a single click. The data covers a wide range of classes that can allow initial exploration of many anatomical landmarks and pathological findings. 

The main contributions of this work are summarized below: 

\begin{itemize}
\item We present an open-access multi-class GI endoscopy dataset, namely, \textit{Gastrovision}, containing 8,000 images with 27 classes from two hospitals in Norway and Sweden. The dataset exhibits a diverse range of classes, including anatomical landmarks, pathological findings, polyp removal cases and normal or regular findings. It covers a wide range of clinical scenarios encountered in endoscopic procedures. 

\item We evaluated a series of deep learning baseline models on standard evaluation metrics using our proposed dataset. With this baseline, we invite the research community to improve our results and develop novel GI endoscopy solutions on our comprehensive set of GI finding classes. Additionally, we encourage computer vision and machine learning researchers to validate their methods on our open-access data for a fair comparison. This can aid in developing state-of-the-art solutions and computer-aided systems for GI disease detection and other general machine learning classification tasks. 

\end{itemize}

\begin{table*}[t]
\centering
\caption{List of the existing datasets within GI endoscopy.}
\resizebox{\linewidth}{!}{
\begin{tabular}{|p{0.35\linewidth}|p{0.25\linewidth}|p{0.35\linewidth}|p{0.2\linewidth}|} 
\hline
\bf Dataset &\bf Data type & \bf Size & \bf Accessibility \\ \hline
Kvasir-SEG~\cite{jha2020kvasir} & Polyps & 1,000 images$^\dag$$^\clubsuit$ &  Public\\ \hline
HyperKvasir~\cite{borgli2020hyperkvasir} & \shortstack{GI findings} & \shortstack{110,079 images \\ \& 374 videos} & Public\\ \hline
Kvasir-Capsule~\cite{smedsrud2021kvasir} & \shortstack{GI findings$^\diamond$} & 4,741,504 images & Public \\ \hline

Kvasir~\cite{pogorelov2017kvasir} & \shortstack{GI findings} & 8,000 images & Public \\ \hline

CVC-ColonDB~\cite{bernal2012towards} & Polyps & 380 images$^\dag$ $^\ddagger$ &  As per request$^\bullet$ \\ \hline
ETIS-Larib Polyp DB~\cite{silva2014toward} & Polyps & 196 images$^\dag$ &  Public \\ \hline			
EDD2020~\cite{ali2020endoscopy,ali2021_endoCV2020} & \shortstack{GI lesions}  & 386 images$^\dag$$^\clubsuit$ & Public\\ \hline 
		
CVC-ClinicDB~\cite{bernal2015wm} & Polyps & 612 images$^\dag$ &  Public\\ \hline 
		
CVC-VideoClinicDB~\cite{bernal2017miccai} & Polyps & 11,954 images$^\dag$ &  As per request \\ \hline

\shortstack{ASU-Mayo polyp\\ database~\cite{tajbakhsh2015automated}} & Polyps & 18,781 images$^\dag$ & As per request$^\bullet$  \\ \hline

KID~\cite{koulaouzidis2017kid} & \begin{tabular}[c]{@{}l@{}}Angiectasia,\\ bleeding,\\ inflammations$^\diamond$
 \end{tabular} & \begin{tabular}[c]{@{}l@{}} $>$ 2500 images,\\ 47 videos \end{tabular}  & Public$^\bullet$\\ \hline

PolypGen~\cite{ali2021polypgen} & Polyps &\shortstack{1,537 images$^\dag$$^\clubsuit$ \\ \& 2,225 video sequence,\\ 4,275 negative frame}& Public\\ \hline

SUN Database \cite{misawa2021development} & Polyps & 158,690 video frames$^\clubsuit$  &  As per request\\ \hline

\textbf{GastroVision (ours)} & GI findings & 8,000 image frames & Public\\ \hline
    
\multicolumn{4}{l}{$^\dag$Segmentation ground truth \hspace{.1cm}\hspace{.05cm}$^\bullet$Not available now $^\ddagger$Contour \hspace{.1cm} $^\diamond$Video capsule endoscopy \hspace{.1cm} $^\clubsuit$ Bounding box information}\\
\end{tabular}}
\label{tab:datasets}
\end{table*}

\section{Related Work}
\label{related work}
Table~\ref{tab:datasets} shows the list of the existing dataset along with data type, their size, and accessibility. It can be observed that most of the existing datasets in the literature are from colonoscopy procedures and consist of polyps still frames or videos. These are mostly used for segmentation tasks. Most of the existing datasets are small in size and do not capture some critical anatomical landmarks or pathological findings. In the earlier GI detection works, the CVC-ClinicDB~\cite{bernal2015wm} and CVC-ColonDB~\cite{bernal2012towards} were widely used. \textbf{CVC-ClinicDB} is developed from 23 colonoscopy video studies acquired with white light. These videos provide 31 video sequences, each containing one polyp, which finally generates 612 images of size 576 $\times$ 768. \textbf{CVC-ColonDB} consists of 300 different images obtained from 15 random cases. Similarly, \textbf{ETIS-Larib Polyp DB}~\cite{silva2014toward} is a colonoscopy dataset consisting of 196 polyp frames and their corresponding segmentation masks. Recently, \textbf{Kvasir-SEG}~\cite{jha2020kvasir} dataset has been introduced that comprises of 1,000 colonoscopy images with segmentation ground truth and bounding box coordinate details. This dataset offers a diverse range of polyp frames, including multiple diminutive polyps, small-sized polyps, regular polyps, sessile or flat polyps collected from varied cohort populations. The dataset is open-access and is one of the most commonly used datasets for automatic polyp segmentation. 

The \textbf{ASU-Mayo Clinic Colonoscopy Video (c) database}~\cite{tajbakhsh2015automated} is a copyrighted dataset and is considered the first largest collection of short and long video sequences. Its training set is composed of 10 positive shots with polyps inside and 10 negative shots with no polyps. The associated test set is provided with 18 different unannotated videos.  \textbf{CVC-VideoClinicDB}~\cite{bernal2017miccai} is extracted from more than 40 long and short video sequences. Its training set comprises 18 different sequences with an approximate segmentation ground truth and Paris classification for each polyp. \textbf{SUN Colonoscopy Video Database} comprises 49,136 polyp frames and 109,554 non-polyp frames. Unlike the datasets described above, this dataset includes pathological classification labels, polyp size, and shape information. It also includes bounding box coordinate details.  The \textbf{PolypGen}~\cite{ali2021polypgen} dataset is an open-access dataset that comprises 1,537 polyp images, 2,225 positive video sequences, and 4,275 negative frames. The dataset is collected from six different centers in Europe and Africa. Altogether the dataset provides 3,762 positive frames and 4,275 negative frames. These still images and video frames are collected from varied populations, endoscopic systems, and surveillance expert in  Norway, France, United Kingdom, Egypt, and Italy and is one of the comprehensive open-access datasets for polyp detection and segmentation.

Apart from the lower GI-related datasets, there are a few datasets that provide combined samples of upper and lower GI findings. For example, \textbf{HyperKvasir}~\cite{borgli2020hyperkvasir} is a multi-class GI endoscopy dataset that covers 23 classes of anatomical landmarks. It contains 110,079 images out of which 10,662 are labeled and 99,417 are unlabeled images. The \textbf{EDD2020} dataset~\cite{ali2020endoscopy,ali2021_endoCV2020} is a collection of five classes and 386 still images with detection and segmentation ground truth. The classes are divided into 160 non-dysplastic Barrett’s, 88 suspicious precancerous lesions, 74 high-grade dysplasia, 53 cancer, and 127 polyps with overall 503 ground truth annotations. The \textbf{Kvasir-Capsule}~\cite{smedsrud2021kvasir} is a video capsule endoscopy dataset comprising 4,741,504 image frames extracted from 117 videos. From the total frames, 4,694,266 are unlabeled, and 47,238 frames are annotated with a bounding box for each of the 14 classes. Similarly, \textbf{KID}~\cite{koulaouzidis2017kid} is a capsule endoscopy dataset with 47 videos and over 2,500 images. The images are annotated for normal, vascular, inflammatory, lymphangiectasias, and polypoid lesions.

The literature review shows that most GI-related datasets focus on a single specific finding, such as colon polyps. Some of the datasets are small in size and have ignored non-lesion frames, which are essential for developing algorithms to be integrated into clinical settings. Additionally, many of these datasets are available on request and require approval from the data providers resulting in further delays. A few datasets like Kvasir, HyperKvasir, Kvasir-Capsule and KID provide multiple GI findings. However, Kvasir-Capsule and KID are video capsule endoscopy datasets. The Kvasir dataset has only eight classes, whereas Hyperkvasir has 23 classes. In contrast, our \textit{GastroVision} dataset has 27 classes and covers more labeled classes of anatomical landmarks, pathological findings, and normal findings. Additionally, we establish baseline results on this dataset for GI disease detection and classification, offering valuable research resources for advancing GI endoscopy studies.

\section{GastroVision}
Here, we provide detailed information about the dataset, acquisition protocol, ethical and privacy aspects of data and suggested metrics.  
\begin{figure*}
    \centering
    \includegraphics[width=0.81\linewidth]{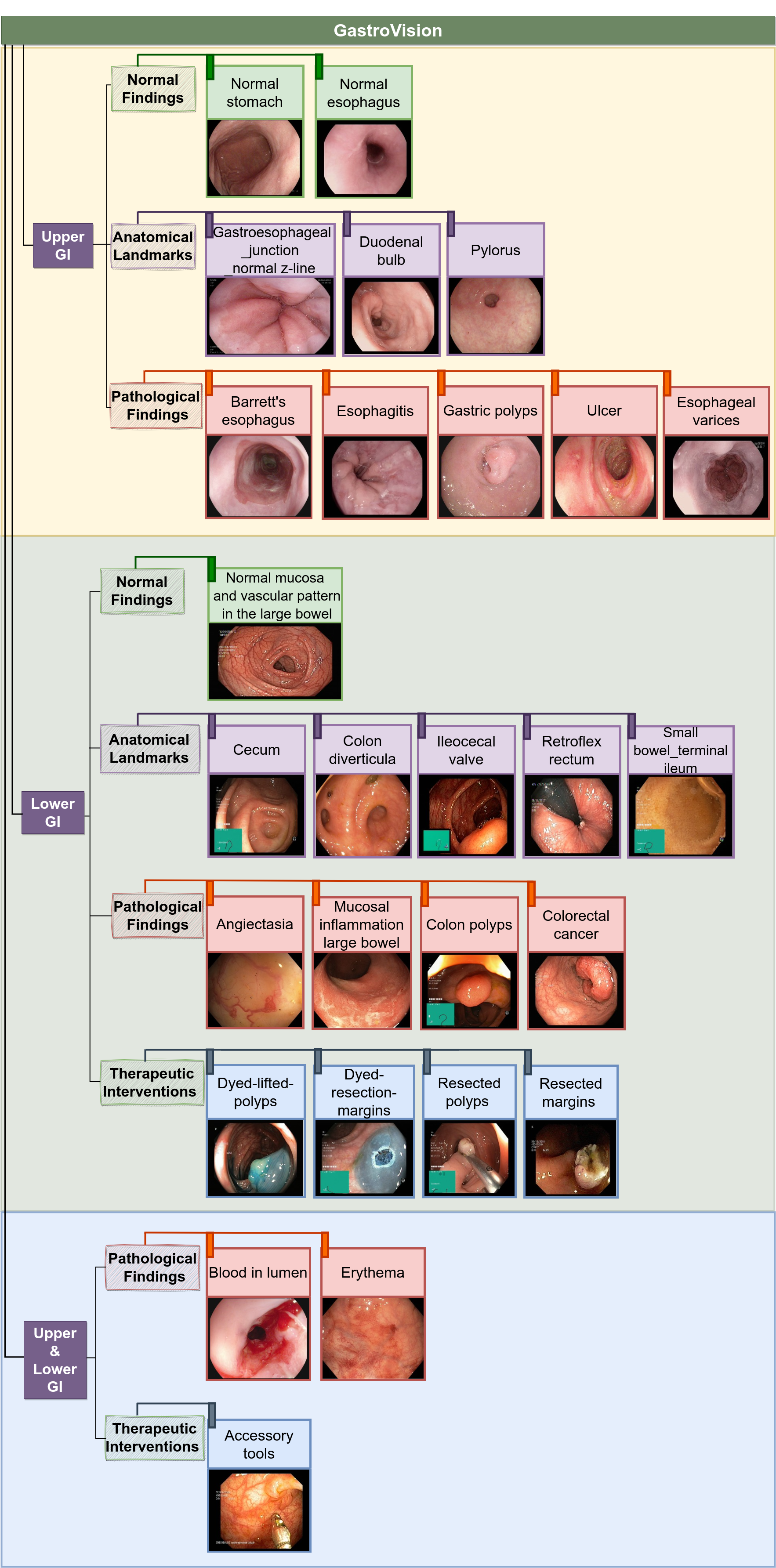}
    \caption{Example images from the gastrointestinal tract showing distinct findings from the upper and lower GI tract.} 
    \label{fig:giclassification}
\end{figure*}

\begin{figure*} [!t]
    \centering
    \includegraphics[width=\linewidth]{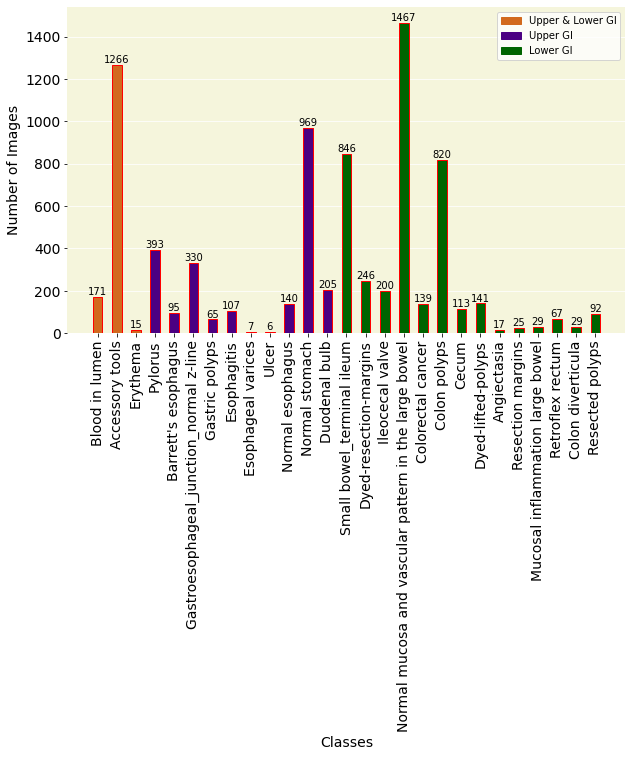}
    \caption{The figure shows the number of images per class. Some classes have few samples because of the rarity of the findings and the technical challenges associated with obtaining such samples in endoscopic settings.}
    \label{fig:datastatistic}
\end{figure*}

\begin{figure} [!t]
    \centering
    \includegraphics[width=0.7\linewidth]{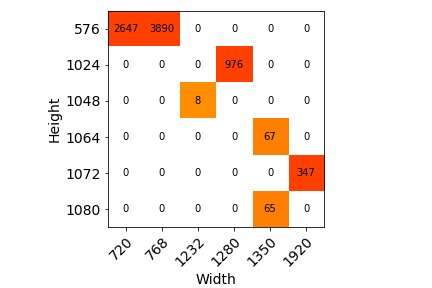}
    \caption{Resolutions of the 8,000 images of GastroVision.}
    \label{fig:resolution}
\end{figure}

\subsection{Dataset details}
GastroVision is an open-access dataset that incorporates 8,000 images pertaining to 27 different labeled classes (Fig.~\ref{fig:giclassification}). Most images are obtained through White Light Imaging (WLI), while a few samples are acquired using Narrow Band Imaging (NBI). These classes are categorized into two broad categories: \textit{Upper GI tract} and \textit{Lower GI tract}. The number of images under each class is presented in Fig.~\ref{fig:datastatistic}. These classes indicate findings acquired from GI tract. It can be observed that the sample count is not balanced across classes, which is generally experienced in the medical image acquisition process as some findings occur less often. 
Releasing these classes in the dataset will allow the researchers to leverage the fast-emerging AI techniques to develop methods for detecting such rare but clinically significant anomalies.  All the images are stored in JPG format, and the overall  size is around 1.8 GB. The resolution details of the images can be found in Fig.~\ref{fig:resolution}. \textit{GastroVision} is provided as a structured directory, with each class having a specific folder. For example, the `\textit{Accessory tools}' folder contains all images featuring diagnostic and therapeutic tools.  

\subsubsection{Upper GI Tract:}
Upper GI endoscopy examines the esophagus, stomach, and duodenum. The various classes covered in this GI tract are discussed below as three subcategories: \textit{normal findings}, \textit{anatomical landmarks}, and \textit{pathological findings}. A detailed categorization is shown in Fig.~\ref{fig:giclassification}. The \textbf{normal stomach} serves as a critical site for initial digestion, while the \textbf{duodenal bulb}, the first part of the small intestine, is critical for nutrient absorption.  Anatomical landmarks are used as reference points to indicate a specific location and assist in navigating during endoscopy procedures. The \textbf{gastroesophageal junction} is an anatomical area where \textbf{esophagus} joins the \textbf{stomach} also alining to the \textbf{normal z-line}, a transitional point where the esophagus's squamous epithelium and the stomach's columnar mucosa lining join. \textbf{Pylorus} is a  sphincter connecting the stomach and the duodenum, the first part of the small intestine.  

Apart from these anatomical landmarks, any pathological conditions may be encountered during endoscopy. \textbf{Esophagitis}, the most common abnormality, is characterized by an inflammation of the esophagus. This disease is graded based on its severity according to the Los Angeles classification. For example, grade B refers to the condition when the mucosal break is limited to the mucosal fold and is more than 5 mm long. In grade D, mucosal break affects 75\% of the esophageal circumference. Long standing {esophagitis} may cause \textbf{Barett's esophagus}, a condition in which the cells of the esophagus's lining start to change, and tissues appear red. This is a precancerous condition. Other frequent lesions observed are \textbf{polyps}, abnormal tissue growth or ulcers.  \textbf{Gastric polyps} are abnormal growths in the stomach lining. \textbf{Ulcers} are the open sores in the stomach or duodenum that can lead to discomfort and bleeding. \textbf{Esophageal varices} result from portal hypertension, causing swollen veins in the esophagus. \textbf{Erythema} refers to redness, often indicating inflammation and \textbf{blood in the lumen} denotes bleeding. \textbf{Accessory tools} aid in investigating and diagnosing upper and lower GI tract conditions for targeted treatment.

\subsubsection{Lower GI Tract:}
The lower GI tract is examined by colonoscopy to investigate any abnormalities in the colon, the rectum, and the terminal ileum (the last part of the small bowel). Here, we covered one more subcategory, \textit{therapeutic interventions}, in addition to \textit{normal findings}, \textit{anatomical landmarks}, and \textit{pathological findings}. A detailed class-wise division is shown in Fig.~\ref{fig:giclassification}. 

The \textbf{normal mucosa and vascular pattern in the large bowel} is essential for absorbing water and electrolytes. The different anatomical landmarks associated with lower GI include \textbf{cecum} (first part of the large intestine), visualizing the appendiceal orifice, \textbf{ileocecal valve} (sphincter muscle between ileum and colon), and the \textbf{small bowl}. During the colonoscopy, these anatomical landmarks act as reference points to prove complete examination. Retroflexion in the rectum is performed to visualize a blind zone, using the bending section of the colonoscope to visualize the distal area of the colon, called \textbf{rectroflex-rectum}. The \textbf{terminal ileum}, the last part of the small intestine, aids in nutrient absorption. \textbf{Colon diverticula}, small pouch-like protrusions, can form along the colon's weakened wall, often in the sigmoid colon~\cite{crafa2022changes}.

During the colonoscopy, the endoscopist navigates through these landmarks and looks for abnormalities such as \textbf{polyps}, \textbf{angiectasia}, and inflammation like \textbf{ulcerative colitis}. \textbf{Angiectasia} is a common lesion representing abnormal blood vessels and is responsible for obscure recurrent lower GI bleeding. These can easily be distinguished from the \textbf{normal vessels} shown in Fig.~\ref{fig:giclassification}. \textbf{Colorectal cancer} occurs in the colon or rectum. One of the early signs of this colorectal cancer can be detected through \textbf{colon polyps}. \textbf{Mucosal inflammation in the large bowel} may be caused by different factors, such as infections or chronic inflammatory conditions.

Apart from the aforementioned pathological conditions, several therapeutic interventions are adopted to treat the detected anomalies effectively. It frequently involves the removal of the lesion/polyp. The surrounding of such \textbf{resected polyps}, also called the \textbf{resection margins} or resection sites, are then considered for biopsies. To enhance lesion demarcation, a solution containing indigo carmine is injected, making resection easier. The appearance of blue color underneath the \textbf{dyed-lifted-polyp} provides accurate polyp margins. After resecting such polyps, the underlying region, known as \textbf{dyed-resection-margin}, appears blue. These margins are important to examine for any remaining tissue of the resected polyp.

\subsection{Dataset acquisition, collection and construction}
\label{sec:data_collection}

\subsubsection{Data acquisition and collection:}
The dataset images are acquired from two centers (Department of Gastroenterology, B{\ae}rum Hospital, Vestre Viken Hospital Trust (VV), Norway and Karolinska University Hospital, Stockholm, Sweden) using standard endoscopy equipment from Olympus (Olympus Europe,
Germany) and Pentax (Pentax Medical Europe, Germany). A team of expert gastroenterologists, one junior doctor, and two computational scientists were involved in the labelling of the images and the related review process. It is worth noting that for dataset collection, we labeled some of the unlabeled images from the HyperKvasir dataset and included them in our dataset. Additionally, we labeled the images acquired from the Karolinska University Hospital to their respective classes for developing a diverse and multi-center ``GastroVision'' dataset.

\subsubsection{Ethical and privacy aspects of the data:}
The dataset is constructed while preserving the patients' anonymity and privacy. All videos and images from B{\ae}rum hospitals were fully anonymized, following the GDPR requirements for full anonymization. Hence, it is exempted from patient consent. The files were renamed using randomly generated filenames. The Norwegian Privacy Data Protection Authority approved this export of anonymized images for research purposes. As the dataset development procedure involved no interference with the medical treatment or care of the patient, it has also been granted an exemption for approval by Regional Committee for Medical and Health Research Ethics - South East Norway. Similarly, the data collection process at Karolinska University Hospital, Sweden, is completely anonymized as per the GDPR requirements.

\subsection{Suggested metrics}

Standard multi-class classification metrics, such as Matthews Correlation Coefficient (MCC), micro and macro averages of recall/sensitivity, precision, and F1-score, can be used to validate the performance using our dataset. MCC provides a balanced measure even in cases with largely varying class sizes. A macro-average will compute the metric independently for each class and then take the average, whereas a micro-average will aggregate the contributions of all classes to compute the metric. Recall presents the ratio of correctly predicted positive observations to all the original observations in the actual class. Precision is the ratio of correctly predicted positive observations to all the positive predicted observations. F1-score integrates both recall and precision and calculates a weighted average/harmonic mean of these two metrics.

\section{Experiments and Results}
In this section, we describe the implementation details, technical validation and the limitation of the dataset. 

\subsection{Implementation Details}
All deep learning diagnostic models are trained on NVIDIA TITAN Xp GPU using PyTorch 1.12.1 framework. A stratified sampling is performed to preserve the similar distribution of each class during 60:20:20 training, validation, and testing split formation. The images are resized to $224 \times 224$ pixels, and simple data augmentations, including random rotation and random horizontal flip, are applied.  All models are configured with similar hyperparameters, and a learning rate of $1e^{-4}$ is initially set with $150$ epochs. An Adam optimizer is used with the \textit{ReduceLROnPlateau} scheduler. More description about the implementation details and dataset can be found on our GitHub page~\footnote{\url{https://github.com/DebeshJha/GastroVision}}. 

\begin{table*} [!t]
\centering
\caption{Results for all classification experiments on the Gastrovision dataset.}\label{table:results}
\scalebox{0.97}{
\begin{tabular}{|r|c|c|c|c|c|c|c|} 
\hline
& \multicolumn{3}{c|}{\textbf{Macro Average}} & \multicolumn{3}{c|}{\textbf{Micro Average}} & \\\hline
\textbf{Method} & \textbf{Prec. } & \textbf{Recall} & \textbf{F1} & \textbf{Prec.} & \textbf{Recall} & \textbf{F1} & \textbf{MCC}\\\hline
ResNet-50~\cite{he2016deep}   & 0.4373 & 0.4379 & 0.4330 & 0.6816 & 0.6816 & 0.6816 & 0.6416 \\ \hline

Pre-trained ResNet-152~\cite{he2016deep}   & 0.5258  & 0.4287 & 0.4496 & 0.6879 & 0.6879 & 0.6879 & 0.6478 \\\hline

Pre-trained EfficientNet-B0~\cite{tan2019efficientnet}  & 0.5285  & 0.4326 & 0.4519 & 0.6759 & 0.6759 & 0.6759 & 0.6351 \\\hline

Pre-trained DenseNet-169~\cite{huang2017densely}  & 0.6075  & 0.4603 & 0.4883 & 0.7055 & 0.7055 & 0.7055 & 0.6685 \\\hline

Pre-trained ResNet-50~\cite{he2016deep}  & 0.6398 & 0.6073 & 0.6176 & 0.8146 &0.8146 & 0.8146 & 0.7921\\ \hline

Pre-trained DenseNet-121~\cite{huang2017densely}  &  \textbf{0.7388} & \textbf{0.6231} & \textbf{0.6504} & \textbf{0.8203} & \textbf{0.8203} & \textbf{0.8203} & \textbf{0.7987}\\ \hline

\end{tabular}
}
\end{table*}

\begin{table}[t]
    \centering
\begin{tabular}{|l|r|r|r|r|}

\hline
\multicolumn{1}{|c|}{\textbf{}}      & \multicolumn{3}{c|}{\textbf{Pre-trained DenseNet-121}}              & \multicolumn{1}{l|}{}                 \\ \hline
\multicolumn{1}{|c|}{\textbf{Class}} & \multicolumn{1}{l|}{\textbf{Precision}} & \multicolumn{1}{l|}{\textbf{Recall}} & \multicolumn{1}{l|}{\textbf{F1-score}} & \multicolumn{1}{l|}{\textbf{Support}} \\ \hline
Accessory tools                                       & 0.93 & 0.96 & 0.95 & 253 \\ \hline
Barrett's esophagus                                   & 0.55 & 0.32 & 0.4  & 19  \\ \hline
Blood in lumen                                        & 0.86 & 0.91 & 0.89 & 34  \\ \hline
Cecum                                                 & 0.33 & 0.17 & 0.23 & 23  \\ \hline
Colon diverticula                                     & 1    & 0.33 & 0.5  & 6   \\ \hline
Colon polyps                                          & 0.78 & 0.87 & 0.82 & 163 \\ \hline
Colorectal cancer                                     & 0.63 & 0.41 & 0.5  & 29  \\ \hline
Duodenal bulb                                         & 0.72 & 0.76 & 0.74 & 41  \\ \hline
Dyed-lifted-polyps                                    & 0.86 & 0.86 & 0.86 & 28  \\ \hline
Dyed-resection-margins                                & 0.94 & 0.92 & 0.93 & 49  \\ \hline
Esophagitis                                           & 0.5  & 0.23 & 0.31 & 22  \\ \hline
Gastric polyps                                        & 0.6  & 0.23 & 0.33 & 13  \\ \hline
Gastroesophageal\_junction\_normal z-line             & 0.65 & 0.85 & 0.74 & 66  \\ \hline
Ileocecal valve                                       & 0.74 & 0.7  & 0.72 & 40  \\ \hline
Mucosal inflammation large bowel                      & 1    & 0.33 & 0.5  & 6   \\ \hline
Normal esophagus                                      & 0.72 & 0.82 & 0.77 & 28  \\ \hline
\begin{tabular}[c]{@{}l@{}}Normal mucosa and vasular pattern in the \\ large bowel \end{tabular} & 0.81 & 0.87 & 0.84 & 293 \\ \hline
Normal stomach                                        & 0.9  & 0.86 & 0.88 & 194 \\ \hline
Pylorus                                               & 0.8  & 0.92 & 0.86 & 78  \\ \hline
Resected polyps                                       & 0.33 & 0.11 & 0.17 & 18  \\ \hline
Retroflex rectum                                      & 0.75 & 0.43 & 0.55 & 14  \\ \hline
Small bowel\_terminal ileum                           & 0.86 & 0.85 & 0.85 & 169 \\ \hline
\end{tabular} 
\caption{Class-wise performance associated with the best outcome obtained using pre-trained DenseNet-121.}
\label{tab:classresults}
\end{table}

\begin{figure}[!t]
    \centering
    \includegraphics[width=\linewidth]{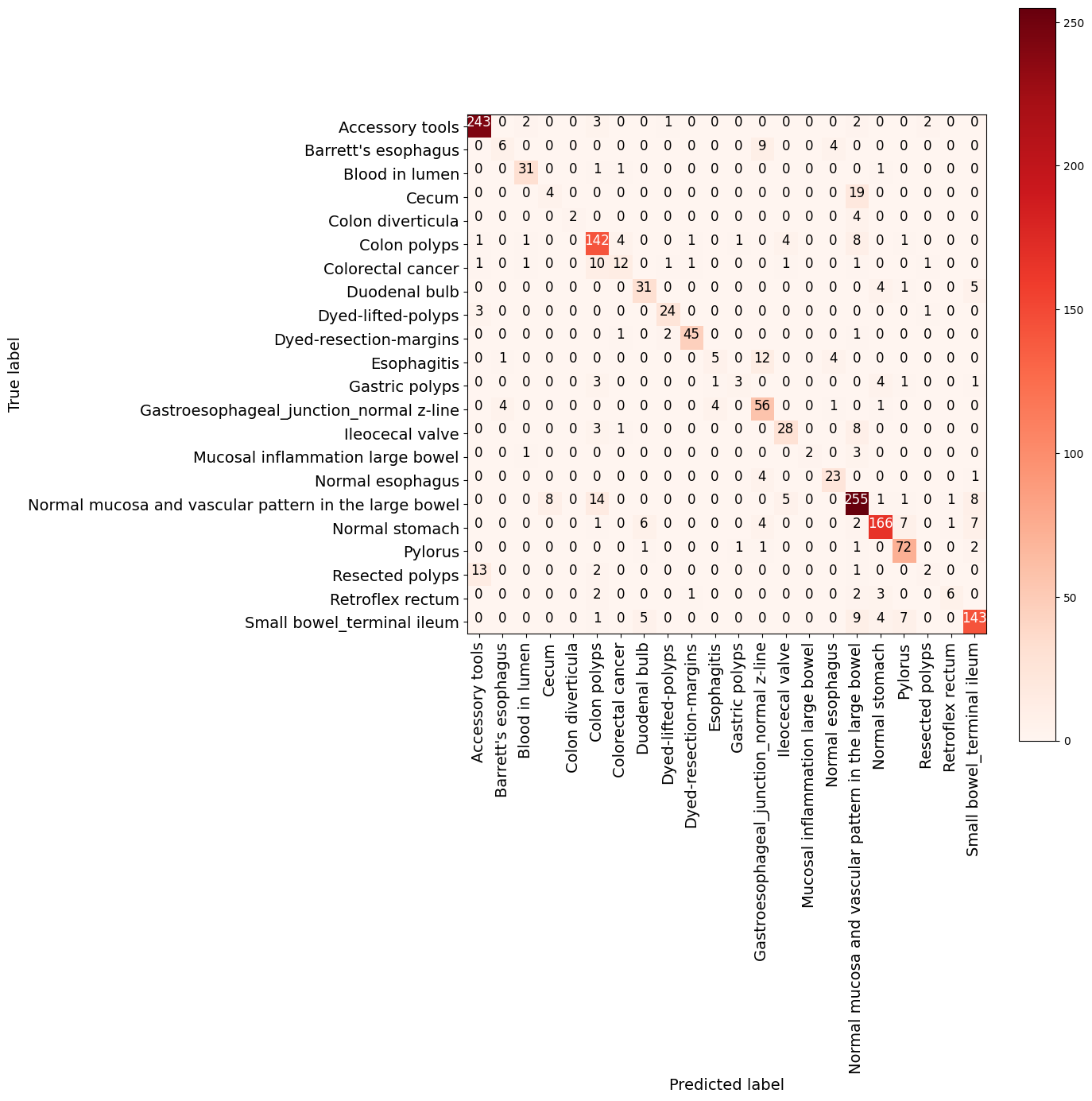}
    \caption{Confusion matrix for the best outcome obtained using pre-trained DenseNet-121.}
    \label{fig:conf_mat}
\end{figure}

\begin{figure*} [!t]
\centering
\includegraphics[width=\linewidth]{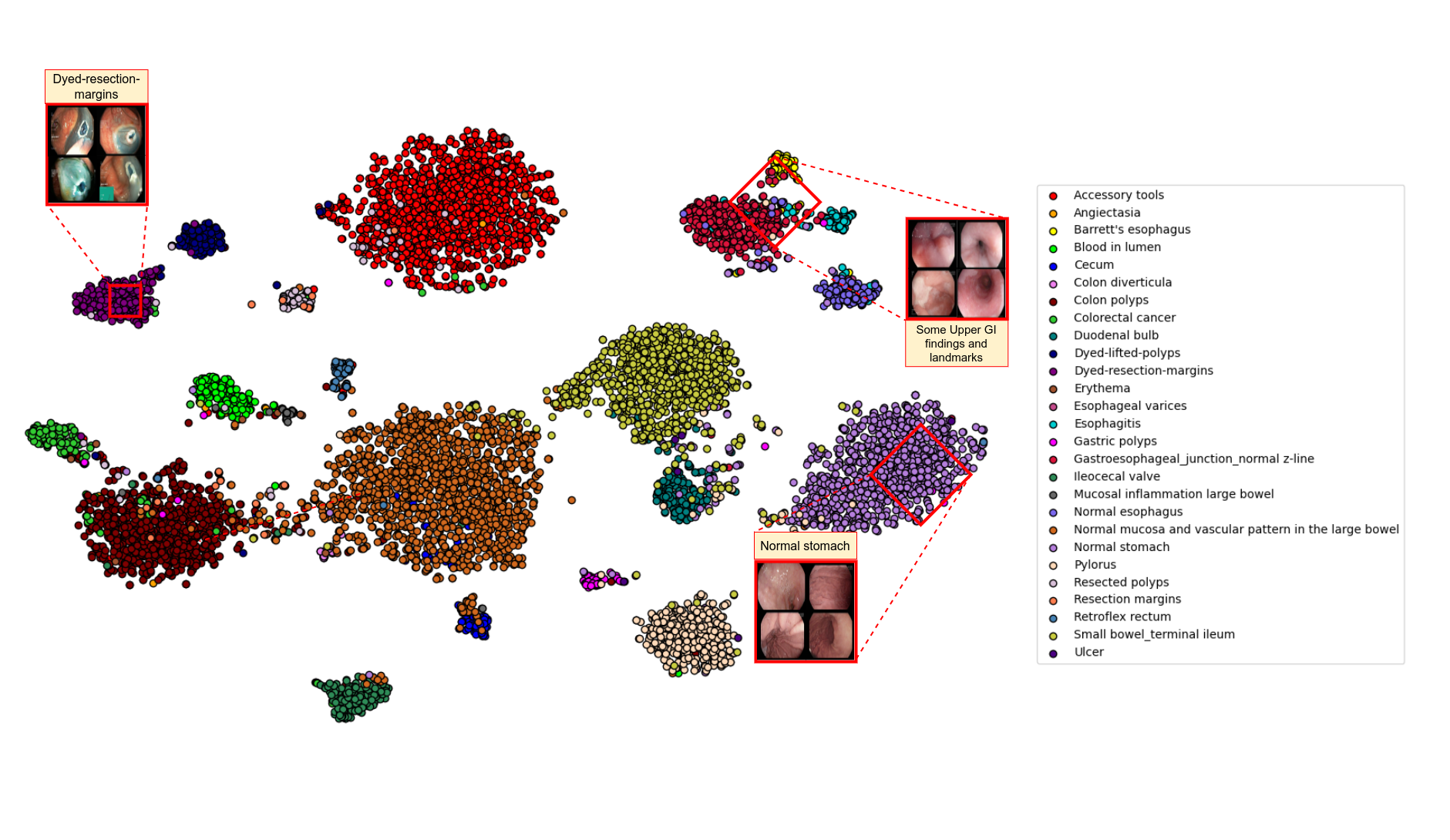}
\caption{Two-dimensional t-SNE embedding for GastroVision. The pre-trained DenseNet-121 model, which is further trained on our training set, is used to extract features. Some sample images are shown with either a specific or a broader (due to multiple overlapping classes) categorization.} 
\label{fig:tnse}
\end{figure*}

\subsection{Technical Validation}
To evaluate the presented data for technical quality and classification tasks, we performed a series of experiments using some state-of-the-art deep learning models. The purpose of this preliminary validation is to provide baseline results that can be referred to for comparison by future researchers. We carried out multi-class classification using CNN-based models, namely, ResNet-50~\cite{he2016deep}, ResNet-152~\cite{he2016deep}, EfficientNet-B0~\cite{tan2019efficientnet}, DenseNet-121~\cite{huang2017densely}, and DenseNet-169~\cite{huang2017densely},  considering their competent performance in GI-related image-based tasks in the literature \cite{thambawita2018medico}. Note that we have only included classes with more than 25 samples in the experiments, which resulted in 22 classes in total. However, we also release the other classes with fewer samples to welcome new interesting findings in areas similar to one-shot learning.

The different experiments performed include \textit{(a) ResNet-50}: The model is randomly initialized, and an end-to-end training is done, \textit{(b) Pre-trained ResNet-50} and \textit{(c) Pre-trained DenseNet-121}: The models are initialized with pre-trained weights, and then all layers are fine-tuned, \textit{(d) Pre-trained ResNet-152}, \textit{(e) Pre-trained EfficientNet-B0} and \textit{(f) Pre-trained DenseNet-169}: The models are initialized with pre-trained weights, and only the updated last layer is fine-tuned. All the above pre-trained models use ImageNet weights. The associated results are shown in Table \ref{table:results}. It can be observed that the best outcome is obtained using the pre-trained DenseNet-121. A class-wise analysis using the same model is provided in Table~\ref{tab:classresults} and Fig.~\ref{fig:conf_mat}. It shows that while most classes achieved satisfactory prediction outcomes, a few proved to be very challenging for the classification model. For a more detailed analysis, we plotted a two-dimensional t-SNE embedding for \textit{GastroVision} (Fig.~\ref{fig:tnse}). 
The classes like \textit{Normal stomach}, \textit{Dyed-resection-margins}, which present a clear distinction in the t-SNE embedding, are less often misclassified. The above points could be the reasons for the F1-score of 0.88 and 0.93 in the case of \textit{Dyed-resection-margins} and \textit{Normal stomach} classes, respectively. On the other hand, there are some overlapping classes such as \textit{Cecum} and \textit{Normal mucosa and vascular pattern in the large bowel} or \textit{Colorectal cancer} and \textit{Colon polyps} which do not present clear demarcation with each other and hence, are likely to be misclassified.

Considering the overall results and many overlapping classes (without distinct clustering), it can be inferred that classifying GI-related anatomical landmarks and pathological findings is very challenging. Many abnormalities are hard to differentiate, and the rarely occurring findings have higher chances of getting misclassified. This presents the challenge of developing a robust AI system that could address multiple aspects important for GI image classification, e.g., many findings are subtle and difficult to be identified, and some findings are not easily acquired during the endoscopy procedure, which results in less number of data samples. Such underrepresented classes need to be explored with some specific algorithms specially designed to leverage the availability of a few hard-to-find samples. Thus, the potential of the baseline results and associated issues and challenges motivate the need to publish this dataset for further investigations.

\subsection{Limitation of the dataset}
Our dataset, \textit{GastroVision}, is a unique and diverse dataset with the potential to explore a wide range of anatomical and pathological findings using automated diagnosis. Although this labeled image data can enable the researchers to develop methods to detect GI-related abnormalities and other landmarks, it lacks segmented annotations in the current version, which could further enhance the treatment experience and surgical procedures. It is important to note that some classes (for example, colon diverticula, erythema,  cecum, esophagitis, esophageal varices, ulcer and pylorus) have only a few images. Despite this limitation, our dataset is well suited for one-shot and few-shot learning approaches to explore some GI-related conditions that have still not received attention in medical image analysis. In the future, we plan to extend the dataset by including more classes and a larger number of samples, along with ground truth for some of the classes that could be used for segmentation purposes as well as images with higher resolution from the most recent endoscopy systems.

\section{Conclusion}
In this paper, we presented a new multi-class endoscopy dataset, \textit{GastroVision}, for GI anomalies and disease detection. We have made the dataset available for the research community along with the implementation details of our method. The labeled image data can allow researchers to formulate methodologies for classifying different GI findings, such as important pathological lesions, endoscopic polyp removal cases, and anatomical landmarks found in the GI tract. We evaluated the dataset using some baseline models and standard multi-class classification metrics. The results motivate the need to investigate better specific techniques for GI-related data. Having a diverse set of categories labeled by expert endoscopists from two different centers, \textit{GastroVision} is unique and valuable for computer-aided GI anomaly and disease detection, patient examinations, and medical training.

\section*{Acknowledgements}
D. Jha is supported by the NIH funding: R01-CA246704 and R01-CA240639.V. Sharma is supported by the INSPIRE fellowship (IF190362), DST, Govt. of India.

\bibliographystyle{splncs04}
\bibliography{mybibliography}

\end{document}